\documentstyle[epsf, rotate]{elsart}

\begin{document}

\begin{frontmatter}

\title{Numerical Determination of the Distribution of Energies for the XY-model}
\author{R. Salazar$^{1}$, R. Toral$^{2}$, A.R. Plastino$^{3}$}
\address{
$^1$ Schuit Institute of Catalysis (SKA), Eindhoven University of Technology,
P.O. Box 513, 5600 MB Eindhoven,  The Netherlands.\\
$^2$Instituto Mediterr\'aneo de Estudios Avanzados (IMEDEA) CSIC-UIB,\\
Campus UIB, E-07071 Palma de Mallorca, Spain .\\
$^3$ CONICET and Faculty of Astronomy and Geophysics, National University
La Plata, Casilla de Correo 727, La Plata 1900, Argentina.
}

\begin{abstract}
We compute numerically the distribution of energies $\Omega(E,N)$ for the XY--model with
short--range and long--range interactions. We find that in both cases the distribution
can be fitted to the functional form: $\Omega(E,N)\sim \exp(N \phi(E,N))$, with $\phi(E,N)$
an intensive function of the energy.
\end{abstract}
\vspace{-0.5cm}

\end{frontmatter}

In 1988 C. Tsallis introduced a generalized $q$-entropy given by
\cite{tsa88}
\begin{equation} \label{eq1}
S_q = \frac { \sum_{i=1}^W p_i^q - 1}{1 - q},
\end{equation}
\vspace{-0.5cm}
\noindent where  $p_i, \,\, (i = 1,\dots,W)$ is the probability of
the microscopic configuration $i$, and the parameter $q$
characterizes the degree of non--extensivity of $S_q$.\footnote{A
quantity is said to be non--extensive if its asymptotic dependence
on the system's size $N$ is non--linear. On the contrary,
extensive quantities exhibit a linear dependence on $N$.}
Non--extensive systems with long--range interparticle interactions
are good candidates to be studied under the generalized
thermostatistics derived from this $q$-entropy \cite{AO01}. One
such system which has received special attention recently
\cite{tam00,cam00,lat01} is the XY model with long--range
interparticle interactions. This model is described by the
Hamiltonian
\begin{equation} \label{eq2}
{\cal H} = \sum_{(i,j)}^N \frac{1-\cos(\theta_i -
\theta_j)}{r_{ij}^{\alpha}},
\end{equation}
\vspace{-0.5cm}
\noindent where $\{\theta_i\}_{i=1,\dots,N}$ are angle--type
variables, and the sum runs over all distinct pairs of sites on a
$d-$dimensional regular lattice of lineal size $L = N^{1/d}$ with
periodic boundary conditions. The distance between the sites $i$
and $j$ is $r_{ij}$, and the parameter $\alpha$ sets the
interaction range. The short--range interaction XY model is
recovered in the limit $\alpha = \infty$. To determine the range
of parameters leading to non-extensive features, it is instructive
to consider the scaling behaviour of the mean energy per particle
\cite{tsa95},
\vspace{-0.2cm}
\begin{equation} \label{eq3}
\frac{E}{N} \sim \tilde N \equiv 1+ d \int_1^L dr r^{d-1}r^{-\alpha}
= \frac {N^{1-\alpha/d}-\alpha/d}{1-\alpha/d}
\end{equation}
\vspace{-0.5cm}
\noindent
 Therefore, for $\alpha>d$ (including the limiting case
$\alpha=\infty$) the energy scales as $E \sim N$ and the system
behaves linearly, whereas in the non--extensive regime $\alpha <
d$, the energy scales as $E \sim N^{2-\alpha/d}$. In the limiting
case $\alpha=d$ the energy per site scales as the logarithm of the
system size: $E\sim N \ln N$. The number of microscopic
configurations with energy between $E$ and $E+\delta E$ is equal
to $\Omega(E,N) \delta E$, where $\Omega(E,N)$ is the density of
states. Our aim here is to compute numerically $\Omega(E,N)$ in
order to study its asymptotic dependence on $N$.

Recent numerical studies of a dynamical version of the long-range
XY model (endowed with appropriate kinetic energy terms) have
shown the existence of metaestable states exhibiting a Tsallis'
maximum entropy distribution of velocities \cite{lat01}. The
lifetime of these metastable states increases with the system
size, thus suggesting that it diverges in the thermodynamic limit.
Unfortunately the relaxation time needed to arrive to these
metastable states also increases with the system size, making it
very difficult to study numerically the proposed
non--commutativity: $\lim_{N \to \infty}\lim_{t \to
\infty}\ne\lim_{t \to \infty}\lim_{N \to \infty}$, which is
expected to appear in the  thermodynamic limit in these kind of
non--extensive systems.

Here we are going to consider an angle of this subject involving a
recently conjectured relation \cite{tsa01} between (i) the range
of applicability of Tsallis Statistics and (ii) the functional
form of the density of states $\Omega(E,N)$. For most systems
studied so far one has $\Omega(E,N) \sim \exp(N \phi(E,N))$, with
$\phi(E,N)$ an intensive function, implying that the natural
expression for an extensive entropy is the well known
Boltzmann--Gibbs entropic form $S = \ln(\Omega(E,N))$. However,
things might be different if, instead, the behaviour of
$\Omega(E,N)$ were as $\Omega(E,N) \sim \varphi(E,N) N^{\mu}$
(again with $\varphi(E,N)$ an intensive function). In such a case
it is conjectured that the relevant entropic form is the one
determined by Eq.({\ref{eq1}), $S_q =(\Omega(E,N)^{1-q}-1)/(1-q)$,
with $\mu(1-q)=1$. Our intention is to present briefly some
results about the form of $\Omega(E,N)$ for the XY model with
long--range interactions. It is shown that an unrestricted
counting of states leads to a functional form $\Omega(E,N) \sim
\exp(N \phi(E))$. Therefore, if ergodicity is satisfied, the
Boltzmann--Gibbs entropic form is the appropriate one. This
suggests, in accord with the results reported in \cite{lat01},
that transient regimes (during which the system is not able to
explore the whole range of energy values) may constitute an
appropriate field of application of the nonextensive
thermostatistics (see also \cite{AO01}).
\begin{figure}[!ht]
\centerline {\epsfxsize=7.7cm \epsfbox{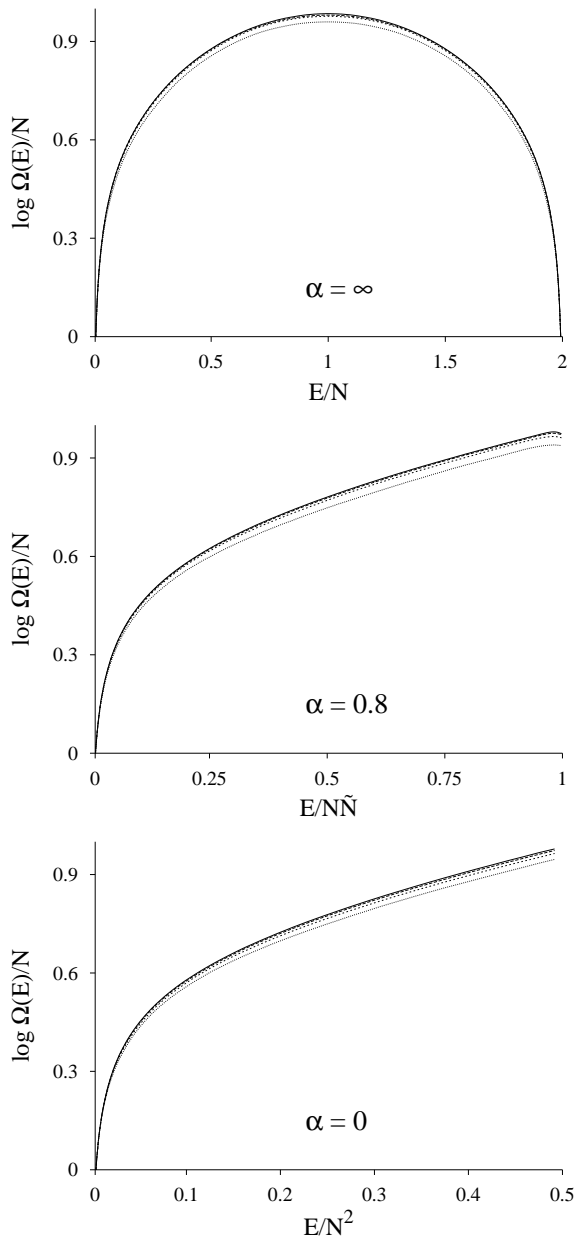}}
\caption{ Scaling relations of the number of states $\Omega(E,N)$
of the one dimensional XY model, Eq.(\ref{eq2}), for different
values of the parameter $\alpha$ and $N = 50, 100, 200, 400$.}
\label{fig}
\end{figure}
The number of states $\Omega(E,N)$ has been numerically computed
using the so--called Histogram by Overlapping Windows (HOW) method
\cite{bha87,sal01,sal99}. Briefly, the HOW method works by
confining the energy histogram generation to a suitably small
energy interval $[E_i,E_i+\Delta E]$, such that the values of the
number of states for each energy in that interval are comparable.
The process is repeated varying $E_i$ until the full energy range
is covered. We have also applied a method recently developed by
Wang and Landau \cite{wan01} which samples the function
$\Omega(E,N)$ autoconsistently. Both methods give the same results
within errors.

We have obtained numerically $\Omega(E,N)$ for the one--dimensional XY model
defined in Eq.({\ref{eq2}), for different system sizes $N = 50, 100, 200, 400$.
Our results are shown in Fig.(\ref{fig})
in three cases: (i) the extensive, short--range limit $\alpha =
\infty$, (ii) the non--extensive case with $\alpha = 0.8$, and
(iii) the non--extensive case with infinite range limit $\alpha =
0$. It is clear from Fig.(\ref{fig}) that in all the cases
considered the generic behavior is $\Omega(E,N) \sim \exp(N
\phi(E,N))$ with an intensive function $\phi(E,N)=\phi(E/N\tilde
N)$ in which the energy appears rescaled according to
Eq.(\ref{eq3}). In this case, and according to the above
conjecture, it appears that the {\sl equilibrium} properties of
this system would be described by the standard canonical ensemble
while the results of \cite{lat01} apply to the metastable states
developed during the transient dynamics.

Finally, we note that the scaling functions $\phi(x)$ appears to be independent
of $\alpha$ in the range $0\le \alpha<d$. This equivalence for this range of
values of $\alpha$ was also observed with other properties of the model in
\cite{tam00,cam00} where it was shown that a mean--field description holds in this case.

We wish to thank C. Tsallis for interesting discussions during the NEXT 2001 conference. We acknowledge financial support from  DGES (Spain) project
PB97-0141-C02-01 and MCyT (Spain) project BMF2000-0624.


\vspace{-0.5cm}
\begin{thebibliography}{00}
\vspace{-0.5cm}

\bibitem{tsa88}
C. Tsallis.
\newblock {\em Journal of Statistical Physics}, 52:479, 1988.
For an updated bibliography on Tsallis Statistics see
http://tsallis.cat.cbpf.br/biblio.htm

\bibitem{AO01}
S. Abe and Y. Okamoto (Eds.) ``Nonextensive Statistical Mechanics
and Its Applications", Springer-Verlag, Berlin, Heidelberg, 2001.

\bibitem{tam00}
F. Tamarit, C. Anteneodo.
\newblock {\em Physical Review Letters}, 84:208, 2000.

\bibitem{cam00}
A. Campa, A. Giansanti, D. Moroni.
\newblock {\em Physical Review E}, 62:303, 2000.

\bibitem{lat01}
V. Latora, A. Rapisarda, C. Tsallis. {\em preprint} cond-mat/0103540.

\bibitem{tsa95}
C. Tsallis
\newblock {\em Fractals}, 3:541, 1995.

\bibitem{tsa01}
C. Tsallis.
Discussion at the {\em NEXT 2001} conference in Cagliari (May 2001).

\bibitem{bha87}
G. Bhanot, R. Salvador, S. Black, P. Carter, R. Toral.
\newblock {\em Physical Review Letters}, 59:803, 1987.

\bibitem{sal01}
R. Salazar, R. Toral.
\newblock {\em Physica A}, 290:159, 2001.

\bibitem{sal99}
R. Salazar, R. Toral.
\newblock {\em Physical Review Letters}, 83:4233, 1999.

\bibitem{wan01}
F. Wang, D. P. Landau.
\newblock {\em Physical Review Letters}, 86:2050, 2001.


\end{thebibliography}
\end{document}